# Superconductivity up to 14.2 K in MnB$_4$ under pressure

Zhe-Ning Xiang[†], Ying-Jie Zhang[†], Qing Lu[†], Qing Li*, Yiwen Li, Tianheng Huang, Yijie Zhu, Yongze Ye, Jian Sun*, and Hai-Hu Wen*

National Laboratory of Solid State Microstructures and Department of Physics, Collaborative Innovation Center of Advanced Microstructures, Nanjing University, Nanjing 210093, China

email: liqing1118@nju.edu.cn, jiansun@nju.edu.cn, hhwen@nju.edu.cn
[†] These authors contributed equally to this work.

*Abstract:* **The discovery of superconductivity in 3*d*-transition metal compounds with strong magnetism is interesting but rare. Especially for Mn-based compounds, there exist only very limited materials that show superconductivity. Here, we report the discovery of superconductivity up to 14.2 K in a Mn-based material MnB$_4$. By applying high pressures, we found the continuous suppression of a weak insulating behavior and the occurrence of superconductivity after about 30 GPa. With further increasing pressure, *T*$_c$ is gradually enhanced and reaches the maximum value of about 14.2 K at 150 GPa with a Fermi-Liquid behavior in the normal states. The synchrotron X-ray diffraction data reveal the unchanged monoclinic (S.G: *P*2$_1$/c) symmetry but an unusual crossover of the lattice parameters *b* and *c*. Theoretical calculations based on the electron-phonon coupling picture reveal a very low *T*$_c$ (less than 1 K), manifesting an exotic pairing mechanism beyond the Bardeen-Cooper-Schrieffer (BCS) theory. Our findings show a promising way to explore high *T*$_c$ superconductivity by combining the 3*d*-transition metal magnetic elements and light elements.**

# Introduction

The discoveries of high-$T_c$ superconductivity in cuprates, [1] iron-based, [2] and nickelate superconductors [3] have stimulated tremendous research interests in the exploration of superconductivity (SC) in 3$d$-transition metal compounds. However, due to the strong incompatibility of magnetism and SC in the electron-phonon based picture, superconductivity is rarely observed in 3$d$ elements with strong magnetism, like Cr or Mn-based compounds. [4-14] Although long-range magnetism is commonly antagonistic to superconductivity, spin fluctuations are thought to be an important ingredient for magnetism driven Cooper pairing in unconventional SC, which occurs in cuprates, iron-based materials and heavy fermion superconductors, *etc.* [15-17] By employing external tuning parameters such as high pressure or chemical doping, the itineracy and localization of $d$-orbital electrons can be effectively modulated [18]. In general, unconventional superconductivity emerges due to strong electronic correlations and in the proximity to magnetic instability when the antiferromagnetic order or bad metal behavior is suppressed, giving a superconducting transition in the vicinity of the possible quantum critical point (QCP) [19, 20].

The first Mn-based SC was discovered in MnP in 2015, by applying pressure to suppress the helimagnetic order, the maximum $T_c$ of about 1 K was observed around 8 GPa. [9] Up to now, only a few Mn-based materials have been found to exhibit superconductivity, such as MnSe ($T_c$ = 9 K at 35 GPa), [10] MnSb$_4$Te$_7$ ($T_c$ = 2.2 K at 50.7 GPa). [11] Recently, pressure-induced unconventional SC is observed in a series of ternary compounds $A$Mn$_6$Bi$_5$ ($A$ = K, Rb, and Cs) with a unique quasi-one-dimensional

(Q1D) structure and the highest $T_c$ = 9.5 K (RbMn$_6$Bi$_5$) is reached among the Mn-based compounds.[12-14] Our recent work on Ti$_{1-x}$Mn$_x$ alloys reveals the bulk superconductivity in beta-phase samples at ambient pressure and a record high $T_c$ up to 26 K is achieved under high pressures. [21, 22] This observation might provide a new platform to explore high-$T_c$ SC in alloys containing rich magnetic element Mn.

Materials with rich light elements are promising candidates for the high-$T_c$ SC, [23, 24] but it can only be achieved by using a high pressure. According to theoretical perspective based on the BCS theory, elements with light mass can provide high Debye frequency, thereby contributing to high critical temperatures. [23, 25] For example, in hydrogen-rich compounds, experiments have shown that several polyhydride compounds can transform into the superconducting state at temperatures above 200 K. [26-28] In these hydrides, the internal chemical compression results in a strong reduction of the metallization pressure, thus the external pressure for achieving superconductivity becomes experimentally accessible. [24] Besides, the discovery of superconductivity in MgB$_2$ at 39 K has been regarded as a typical example of generating high-$T_c$ SC in light element-based compounds at ambient pressures. [29] In recent years, superconductivity with relatively high $T_c$ was also reported in some metal borides under high pressures, [30-32] indicating the promising potential to achieve SC in boron-based materials with higher transition temperatures.

In this work, by combining the strong correlation effect of 3$d$-transition metals Mn and high Debye frequency provided by light element boron, we report the pressure-induced superconductivity up to 14.2 K in the Mn-based compound MnB$_4$, this $T_c$ value is unprecedentedly high in stoichiometric Mn-based compound. At ambient pressures, $MnB_4$ shows a weak insulating behavior with relatively high Debye temperature as evident from resistivity and specific heat measurements. As pressure increases, the weak insulating behavior of resistance at low temperatures is gradually suppressed and a superconducting transition is observed above 2 K under a pressure of about 30 GPa. Upon further compression, the $T_c$ is gradually enhanced and reaches the maximum value (14.2 K) at about 150 GPa. High-pressure X-ray diffraction data reveals a continuous shrinkage of the cell volume and an unusual crossover of lattice constants $b$ and $c$, which is consistently supported by the theoretical calculations. A possible unconventional pairing mechanism is also proposed in $MnB_4$ since the calculated $T_c$ based on the electron-phonon coupling is much lower than what we observed experimentally.

**Results**

Transition metal tetraborides $T_rB_4$ ($T_r$ = Cr, Mn, and Fe) were reported about 50 years ago and considered as potential superhard materials.[33-36] The super-high hardness and unexpected superconductivity of iron tetraboride were reported in previous work.[37] The basic structure of transition metal tetraborides can be seen as a three-dimensional boron network with $T_r$ atoms located in the center as shown in Fig 1(a). Among all these compounds, $MnB_4$ was reported to be unique with a monoclinic (S.G: $P2_1/c$) crystal structure,[38, 39] in contrast to the orthorhombic (S.G: *Pnnm*) symmetry reported in $CrB_4$ and $FeB_4$.[39, 40] The lower symmetry in $MnB_4$ is caused by the Peierls distortion of the Mn chains along the *a*-axis (right upper panel of Fig. 1(a)). Note that the boron networks can be seen as $B_{12}$ cages and each Mn atom is surrounded by 12 boron atoms, hence forming a $MnB_{12}$ polyhedra (right lower panel of Fig. 1(a)). We synthesize the $MnB_4$ samples, in both polycrystalline and single-crystal form (See Methods and Supplementary Note 1), and present the powder X-ray diffraction data in Fig. 1(b). We can see that all the diffraction peaks can be well indexed (red curve) to the monoclinic structure with the cell parameters $a$ = 5.8963(2) Å, $b$ = 5.3665(2) Å, $c$ = 5.5029(2) Å, and $β$ = 122.704(4) °, respectively. The obtained values are in good agreement with the previous reports. [38]

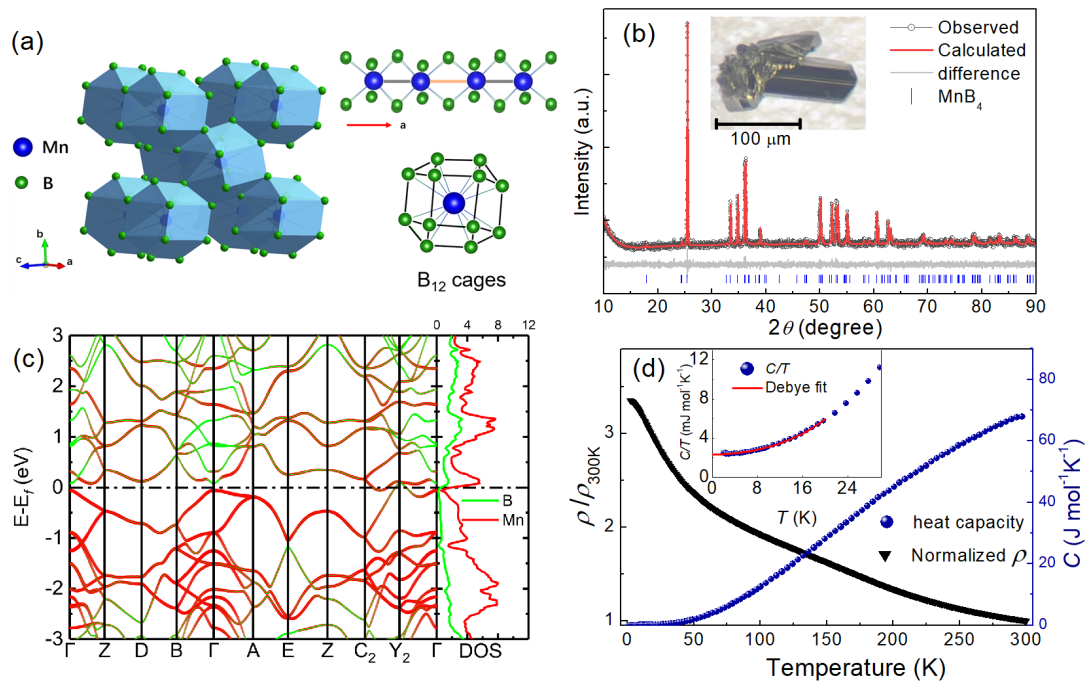

**Fig. 1 | Structure and physical properties of MnB$_4$ at ambient pressure.** (a) The crystal structure of MnB$_4$. The blue and green spheres represent the Mn and B atoms, respectively. Right upper panel: The dimerized Mn-Mn bonds along *a*-axis. There are two kinds of Mn-Mn distance in this direction, namely 2.7 Å for the shorter bond (gray color) and 3.2 Å for the longer bond (orange color). Right lower panel: B$_{12}$ cages around Mn atoms. (b) Powder x-ray diffraction pattern (circles) and Rietveld fitting curve (red line) of MnB$_4$. The inset shows the photograph of a prepared single crystal. (c) Calculated electronic band structures and density of states of MnB$_4$ at ambient pressure.

The red and green colors represent the electronic bands contributed by Mn and B, respectively. A pseudogap feature is observed around the Fermi Level ($E_F$). (d) The temperature-dependent normalized resistivity and specific heat data of MnB$_4$ from 2 to 300 K at ambient pressure. The inset shows the temperature-dependent specific heat coefficient $C/T$ in the temperature region from 2 to 30 K and the fitting to Debye model (red curve).

Due to the Peierls instability, a distortion occurs between two neighboring Mn atoms along the Mn-Mn chain, thus, a gap opening near the Fermi level might be expected. [38, 40, 41] Compared to the undistorted high symmetry structure (orthorhombic, *Pnnm*), the orbital degeneracy near the Fermi level is strongly enhanced due to the dimerization, which shifts the DOS maximum away from Fermi level. [38, 39] Figure 1 (c) displays the electronic structure and density of states at ambient pressure. It is seen that the 3*d*-orbitals of Mn contribute more density of states around the Fermi level and a pseudo-gap feature around the Fermi level is observed due to the Mn-Mn interactions. We measured the temperature-dependent normalized resistivity from 2-300 K of MnB$_4$ and presented the data in Fig. 1(d). The electrical transport shows a semiconducting-like behavior, which is consistent with the calculated band structures and previous results. [38, 39, 42]

Figure 1(d) shows the temperature-dependent heat capacity from 2 to 300 K, which exhibits a monotonic decrease with temperature and. Hence, we present the $C/T$ versus $T$ curve in the low-temperature region and the corresponding fitting curve based on the Debye model, $C/T = \gamma_0 + \beta T^2 + \delta T^4 \ldots$ is shown in the inset of Fig. 1(d). Here, $\gamma_0$ is the specific heat coefficient at zero temperature, $\beta$ and $\delta$ are the temperature-independent fitting parameters. The Debye fitting yields the values of $\gamma_0$ and $\beta$ are 2.35 mJ/(mol-K$^2$) and 0.00739 mJ/(mol-K$^4$), respectively. The small residual term of specific heat ($\gamma_0$) indicates the presence of finite DOS near Fermi energy, which is in contrast to the insulating behavior of resistivity. Meanwhile, the small value of the specific heat coefficient $\gamma_0$ in the zero temperature limit may reflect the existence of the pseudogap feature due to the dimerization of Mn atoms. Using the equation $\theta_D = (12\pi^4 k_B N_A Z / 5\beta)^{1/3}$, where $k_B$ is Boltzmann constant, $N_A$ is Avogadro constant, $Z$ is the number of atoms in one unit cell, we can estimate the Debye temperature ($\theta_D$) of MnB$_4$ at ambient temperature. The value of $\theta_D$ is about 1095 K, which is significantly higher than that of other 3$d$-transition metal compounds,[21, 43] and suggests that the light element B indeed induces a high Debye frequency in MnB$_4$.

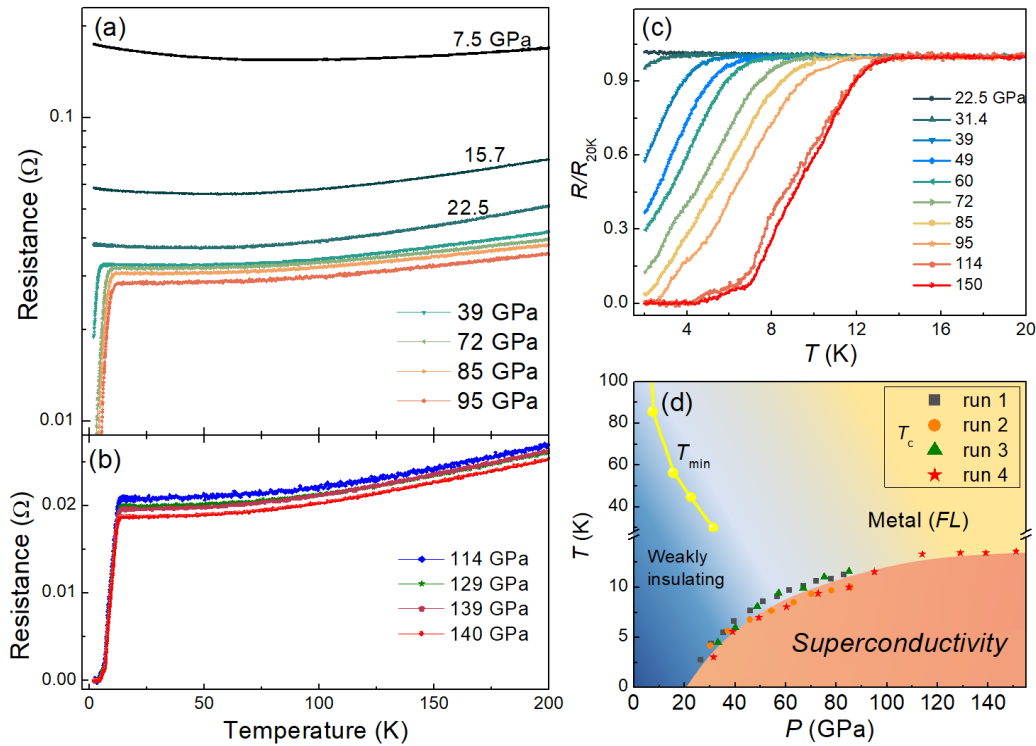

**Fig. 2 | Pressure-induced superconductivity in MnB$_4$ under high pressures.** (a, b) Temperature-dependent resistance (R-T) of MnB$_4$ under various pressures up to 150 GPa. (c) Normalized R/R$_{20K}$-T curves of MnB$_4$ at selected pressures. The evolution of

superconducting transition can be clearly seen. (d) Temperature-pressure phase diagram of MnB$_4$. The solid yellow circles represent the temperature for the upturn of resistance in low temperature region. Below this temperature, the resistance shows a weak insulating behavior. The other color symbols represent the values of $T_c$ from different runs of experiments. Here, $T_c$ represents the onset transition temperature. *FL*: Fermi liquid behavior.

To explore whether superconductivity can be achieved in MnB$_4$ under higher pressure, we performed high-pressure resistance measurements on MnB$_4$ using a diamond anvil cell (DAC) apparatus. Figures 2 (a) and (b) show the temperature-dependent resistance of MnB$_4$ under various pressures up to 150 GPa in a wide temperature range from 2 to 200 K. As we can see, the overall resistance gradually decreases with increasing pressure. Under low pressures, the R-T curves show metallic behavior in high-temperature regions and a resistance upturn in low-temperature regions. Here, we define $T_{min}$ as the temperature corresponding to the minimum value of resistance in low-temperature region. With increasing pressure, the $T_{min}$ gradually reduces to a lower temperature. At about 39 GPa, a sudden drop of resistance is observed in the R-T curve at low temperatures, indicating the occurrence of superconductivity in MnB$_4$. Under higher pressures, the superconducting temperature is gradually enhanced with increasing pressure.

To reveal the complete evolution of the superconducting state under higher pressures, we display the low-temperature R-T curves (normalized at 20 K) in Fig. 2(c). Superconductivity begins to emerge when the pressure reaches 31.4 GPa. After that, the superconducting transition becomes steeper with increasing pressure, and the zero-resistance state is reached above 2 K at about 95 GPa. The highest onset transition temperature reached in our present study is about 14.2 K at 150 GPa as shown in Fig. 2(c) and Fig. S1 (Supporting Information). A zero-resistance state is achieved at about 4 K. It is worth mentioning that the transport behavior above $T_c$ shows a typical Fermi liquid behavior with the exponent $n$ close to 2, see Fig. S1 (Supporting Information). To check the reproducibility and demonstrate the evolution of $T_c$ of $MnB_4$ more clearly, we have carried out additional measurements of resistance under high pressures on three other samples, the results are presented in Fig. S2 (Supporting Information). Superconductivity appears at high pressure in all runs of measurements, and a similar evolution process is observed in all measurements, indicating the good consistency of repeating superconductivity in $MnB_4$. Furthermore, we have also measured the transverse resistivity of $MnB_4$, it is clear that the Hall coefficient $R_H$ is negative and has a moderate temperature dependence (see Fig. S3, Supporting Information).

Based on above experimental observations, we can construct the Temperature-Pressure ($T$-$P$) phase diagram of $MnB_4$ as shown in Fig. 2(d). With the continuous decrease of $T_{min}$, the weak insulating behavior is suppressed progressively. After about 30 GPa, superconductivity starts to appear above 2 K and gets gradually enhanced with increasing pressure, resulting in a half-dome-like superconducting region in the $T$-$P$ phase diagram. At the highest pressure applied in the present work, the onset transition temperature increases to about 14.2 K, the record value among the Mn-based stoichiometric compounds. Interestingly, the $T_c$ value still goes up slowly without saturation after 150 GPa. The slight variation in $T_c$ for different experimental runs may be due to the non-hydrostatic conditions in the high-pressure experiments.

To further characterize the superconducting state of MnB$_4$ under high pressures, we measured the low-temperature R-T curve under various magnetic fields up to 9 T at 114 and 150 GPa, as shown in Figs. 3 (a) and (b). One can see that the superconducting transition gradually decreases as the external magnetic field increases and persists at the highest magnetic field (9 T) in the present study. We also performed the magnetoresistance measurements and *R-T* curves with different electric currents, see Fig. S4 (Supporting Information). These results illustrate that the application of either magnetic fields or electric currents can gradually suppress the superconducting transition.

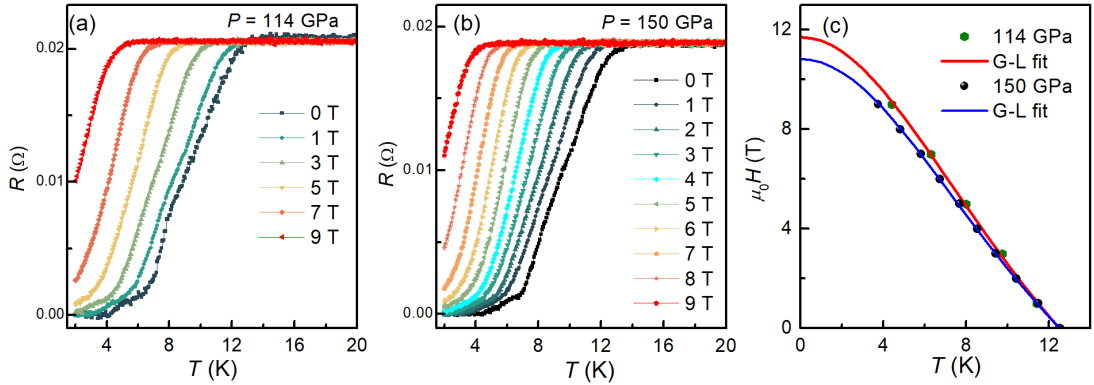

**Fig. 3 | Upper critical fields of MnB$_4$ under high pressures.** (a, b) Temperature-dependent resistance of MnB$_4$ under various magnetic fields at 114 and 150 GPa. (c) The temperature-dependent upper critical field $\mu_0 H_{c2}$ at different pressures. The $T_c$ is determined by using the criterion of temperature for reaching 95% of the normal state resistance. The red and blue lines represent the fitting curve by using the Ginzburg–Landau (G–L) equation.

Based on the R-T curves under different magnetic fields, we can derive the upper critical field $\mu_0 H_{c2}$ as a function of temperature *T*, as plotted in Fig. 3(c). The $\mu_0 H_{c2}$-*T*

data can be well-fitted by using the empirical Ginzburg-Landau formula

$$\mu_0 H_{c2}(T) = \mu_0 H_{c2}(0)[1 - (T/T_c)^2]/[1 + (T/T_c)^2]. \qquad (1)$$

The estimated upper critical field at zero temperature $\mu_0 H_{c2}$ (0) of MnB$_4$ at 114 GPa is 11.7 T, which is lower than its corresponding Pauli paramagnetic limit. We can also calculate the Ginzburg-Landau coherence length $\xi_{GL}$ by using the formula $\mu_0 H_{c2}(0) = \Phi_0/2\pi\xi_{GL}^2$, where $\Phi_0 = h/2e$ is the flux quantum. The calculated $\xi_{GL}$ for MnB$_4$ at 114 GPa is 5.3 nm. As the pressure is increased to 150 GPa, the value of $\mu_0 H_{c2}(0)$ slightly decreases, as shown by the blue fitted line in Fig. 3(c). Moreover, we have also measured the upper critical fields in other experimental runs with different pressures, and the results are shown in Fig. S2(g-i) (Supporting Information). The $\mu_0 H_{c2}$ (0) is roughly consistent with each other and gives an approximate value of 12 T. Moreover, we also extracted the $T_c$ values and the ratios of $\mu_0 H_{c2}(0)/T_c$ of other superconductors containing light element boron and compared them with MnB$_4$, as listed in Table S1 (Supporting Information). To our surprise, the ratio of $\mu_0 H_{c2}(0)/T_c$ in MnB$_4$ has the highest value (0.98) among the reported superconductors containing boron, suggesting different origins. This may indicate that the magnetism contributed by the 3*d* electrons of Mn plays an important role in the pairing, leading to a relatively high upper critical field.

To find out the underlying evolution of crystal structure during the development of superconductivity under high pressures, we performed high-pressure X-ray diffraction (HP-XRD) measurements on MnB$_4$ at the BL15U1 beamline of the Shanghai Synchrotron Radiation Facility. As shown in Fig. 4(a), at a low pressure (5 GPa), almost all diffraction peaks could be well indexed by a monoclinic crystal structure with space group $P2_1/c$. The peak indicated by the asterisk symbol at 18.4° belongs to the gasket Re. With increasing pressure, all reflections shift towards higher angles continuously without emergence of new peaks, indicating a continuous compression of the lattice parameters and the absence of structural phase transition up to 128 GPa. The relative intensity of diffraction peaks experiences some systematic evolution with increasing pressure, which may be due to the inhomogeneity of pressure or the reorientation of grains upon compression as often observed in synchrotron XRD measurements. [44, 45] We also conducted an additional measurement for another sample at higher pressures, the X-ray diffraction patterns and the corresponding Rietveld fitting curves of MnB$_4$ are presented in Fig. S5 (Supporting Information). These fitting results indicate that the MnB$_4$ sample holds the same structure as the one at ambient-pressure all the way up to 158 GPa, the maximum pressure investigated in present study.

The pressure-dependent lattice parameters and cell volume are displayed in Figs. 4(b) and (c). All these parameters monotonously decrease with pressure, indicating the continuous compression of the cell volume under high pressures. The pressure-dependent cell volume of MnB$_4$ sample can be fitted by the third-order Birch-Murnaghan (B-M) equation, [46]

$$P(V) = \frac{3}{2}B_0[(\frac{V_0}{V})^{7/3} - (\frac{V_0}{V})^{5/3}] \times \{1 + \frac{3}{4}(B'_0 - 4)[(\frac{V_0}{V})^{2/3} - 1]\}. \quad (2)$$

Where $V_0$, $B_0$, and $B'$ are the unit cell volume at zero pressure, bulk modulus, and first-order derivative of the bulk modulus. Our fitting yields $V_0 = 147.1$ Å$^3$, $B_0 = 343.05$ GPa, and $B' = 3.39$, respectively. The good consistency between the experimental data and the fitted curves indicates that there is no structural phase transition occurring within the pressure range we measured. We also calculated the enthalpy curves of MnB$_4$ in different type structures and found that the *P*2$_1$/*c*-type structure has the lowest energy (see Fig. S6, Supporting Information), in good agreement with our experimental results. Furthermore, the relatively large bulk modulus indicates that the material has the potential to be a candidate for hard material, as discussed in the previous works.[36, 39]

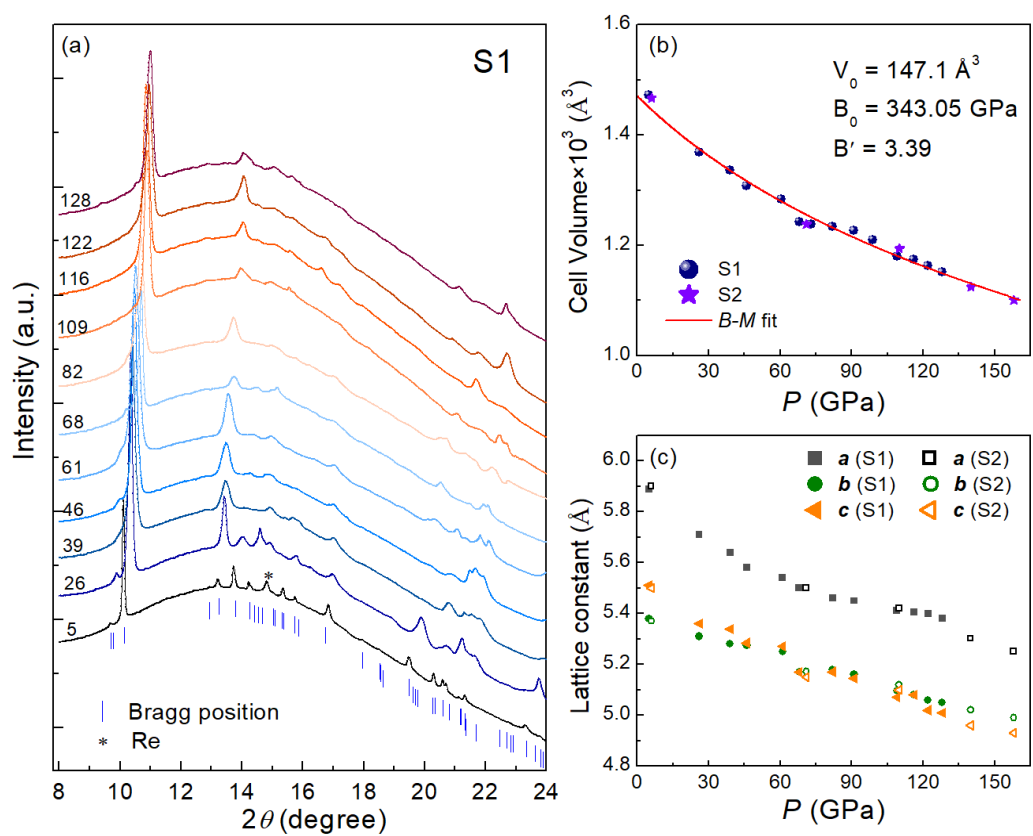

**Fig. 4 | High-pressure synchrotron X-ray diffraction data.** (a) X-ray diffraction patterns of MnB$_4$ collected at different pressures up to 128 GPa. (b) The derived cell volume as a function of pressure for MnB$_4$. The solid red line is the third-order Birch-Murnaghan (B-M) fitting curve. (c) Pressure dependence of lattice parameters of MnB$_4$ obtained from different experimental runs. A crossover of the lattice constant *b* and *c* was observed at around 60-90 GPa.

Figure 4 (c) displays the pressure-dependent lattice parameters extracted from different experimental runs. It is interesting to find that the compression rate of lattice constant $c$ is larger than $b$ under high pressures. Along with the occurrence of superconductivity, the lattice parameters of $b$ and c tend to be the same value at about 60-90 GPa, and $b$ is larger than $c$ in the higher pressure region when the superconductivity is established. Such an unusual behavior can also be supported by theoretical calculation as shown in Fig. S6 (b) (Supporting Information). Considering the transport results under high pressures in Fig. 2, the unusual crossover of the lattice parameters $b$ and $c$ during compression may be closely related to the appearance of superconductivity in MnB$_4$.

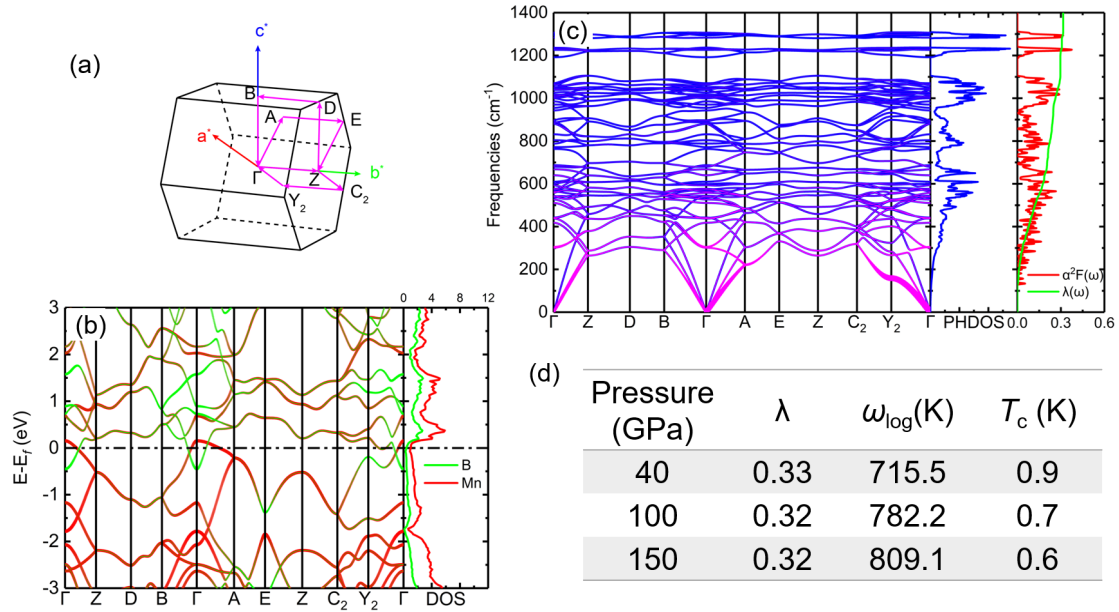

**Fig. 5 | DFT calculations.** (a) The Brillouin zone and high-symmetry-point paths for calculation. (b) Calculated electronic band structures and density of states of MnB$_4$ at 150 GPa. (c) Phonon spectrum, phonon density of states, Eliashberg function $\alpha^2F(\omega)$, and accumulated electron-phonon coupling strength $\lambda(\omega)$ at 150 GPa. (d) Superconducting properties at 40, 100, and 150 GPa, including electron-phonon coupling strength $\lambda$, logarithmic average phonon frequency $\omega_{\log}$ (K), and transition temperature $T_c$ (K).

To further understand the superconductivity with a rather high $T_c$ observed in MnB$_4$

under high pressures, we performed the density functional theory (DFT) calculations on electronic and superconducting properties at varying pressures. Figure 5(a) illustrates the calculated Brillouin zone and the selected high-symmetry-point path for the electronic band and phonon dispersion calculations. The band structure and density of states of MnB$_4$ at 150 GPa show a typical metallic behavior, as illustrated in Fig. 5(b), which is consistent with experiment results. Specifically, compared with the case at ambient pressure in Fig. 1(c), the presence of pseudogap triggers a weak insulating behavior, the pressure-induced crossover of energy bands around Γ point at 150 GPa annihilates the pseudogap and results in a metallic state. The $d$-orbital electrons of Mn atoms occupy the most states at the Fermi level. We further focused on the superconducting properties, with the results shown in Figs. 5(c) and (d), and Fig. S7 (Supporting Information). Although the value of $\omega_{\log}$ increases at higher pressure due to phonon hardening, the calculated $T_c$ values are still fairly small (less than 1 K) at different pressures, predominantly due to the small value of $\lambda$. For example, at 150 GPa, the coupling strength between electrons and phonons below 400 cm$^{-1}$ is the strongest as projected on the phonon spectrum. However, due to the lower phonon density of states below 400 cm$^{-1}$, the total contributions to $\alpha^2 F(\omega)$ and $\lambda(\omega)$ are rather small. At higher frequencies, the phonon density of states is larger, while the projected coupling strength is weaker. Therefore, the value of total $\lambda$ is not large enough to induce high $T_c$. The inconsistency between calculations and experiments indicates that MnB$_4$ is most likely an unconventional superconductor that cannot be explained by a single electron-phonon coupling mechanism.

**Discussion and perspectives**

To confirm the pressure-induced superconductivity and check whether it is an intrinsic property of the title compound, we performed measurements in four experimental runs and all the data show good consistency with each other, which is shown in Fig. 2 and Fig. S2 (Supporting Information). Furthermore, based on the binary Mn-B equilibrium phase diagram, [47] we performed additional high-pressure measurements on potential impurity phases, such as $MnB_2$, $Mn_3B_4$, $MnB$, and $Mn_2B$. As shown in Fig. S8, all the data shows good metallic behavior, and no signature of superconductivity above 2 K is detected. From the high-pressure transport and XRD data, we found that the pressure-induced superconductivity is closely related to the unusual change in lattice parameters. When the lattice parameter *b* exceeds *c*, superconductivity begins to show up and reaches zero resistance as evidenced by the R-T curve in Fig. 2(c). It is commonly observed that the unusual microstructural change will lead to modifications of the electronic structures and generate emergent phenomena, such as superconductivity. [45, 48]

We argue that the superconductivity in $MnB_4$ may have an unconventional origin which is beyond the prediction by the BCS theory. The direct evidence is that the calculated $T_c$ based on electron-phonon coupling is rather low, which is 0.6 to 0.9 K, much lower than what we have observed experimentally. Combined with the band structure calculation results under high pressure and the relatively high $\mu_0 H_{c2}(0)/T_c$ ratio, the 3*d* electrons of Mn definitely play an important role for the superconductivity in MnB$_4$ under high pressure. The 3$d$ electrons can bring about strong correlation effect and induce unconventional superconductivity in the compounds. [3, 5, 12] In addition, due to the special crystal field of MnB$_{12}$ polyhedra (Fig. 1), the 2$p$ electrons of boron may act as a disruptor of the Mn-Mn super-exchange interaction and hence destroy the possible long range magnetic order. This is accomplished by the high-frequency vibration due to the high Debye frequency provided by the light element. In such a case, the magnetic moment given by the 3$d$-transition element, like Mn here provides the basic source for the spin fluctuation which may play a crucial role in the formation of Cooper pairs. It is interesting to note that, in the explorations of high-$T_c$ superconductivity in rich hydrogen materials, people are trying to avoid using the transition metal elements because of their possible magnetic moments. These magnetic moments are harmful for superconductivity formed in the frame of electron-phonon coupling. However, as we addressed above, considering the pairing mechanism in a different way, that is via the magnetic spin fluctuations, then SC with higher $T_c$ may be formed in compounds with 3$d$-transition elements and light elements, like MnB$_4$ here. The light elements will serve as ideal bridges between the magnetic moments and hinder the formation of long-range magnetic order due to its high frequency vibration. This picture relies on the cooperation between magnetic moments of 3$d$-transitions metal elements and high-frequency vibrations of light elements; thus, we abbreviate it as the MHV model. A well-balanced situation between the dual effects of 3$d$-transition metal elements, namely the localization and itinerancy, will lead to the formation of superconductivity. This is certainly an interesting idea, which will open a new avenue for exploring new high-$T_c$ superconductors with combinations of 3$d$-transition metal elements and light elements.

## Conclusion

In summary, by applying high pressures, we discover superconductivity in MnB$_4$ with the $T_c$ up to 14.2 K, which is the highest value among the reported Mn-based stoichiometric superconductors. The establishment of superconductivity was found to be closely related to the abnormal crossover of the lattice parameter $b$ and $c$ under pressures. Under magnetic fields, the ratio of $\mu_0H_{c2}(0)/T_c$ has the maximum value among the superconductors containing boron, suggesting that the 3$d$ electrons of Mn play an important role in the superconductivity in MnB$_4$. More importantly, the superconductivity in MnB$_4$ may have unconventional pairing mechanism since the theoretical calculations based on the picture of electron-phonon coupling give a $T_c$ less than 1 K. Our findings on MnB$_4$ sheds new light on the exploration of potential unconventional superconductivity in compounds with 3$d$-transition metal elements with strong magnetism and light elements. We name this frame of new recipe as MHV model (magnetic moments plus high frequency vibrations).

## Methods

**Sample synthesis and characterization.** The single-crystal and polycrystal samples were both synthesized by the solid-state reaction method. [38, 42] The details about the synthesis process are described in Supplementary Note 1 (Supporting Information). The crystal structure and phase purity were checked by using powder X-ray diffraction (XRD) with Cu $K_\alpha$ radiation (Bruker D8 Advance diffractometer; $\lambda = 1.541$ Å). Rietveld refinement fitting was performed using the TOPAS 4.2 software. [49] The VESTA was used to visualize the crystal structures. The electrical-resistivity and heat capacity measurements were performed on a physical property measurement system (PPMS-9T, Quantum Design). The electrical resistance was obtained by the conventional four-probe method and the specific heat was measured with the thermal-relaxation method.

**High-pressure resistance and *in situ* XRD measurements.** High-pressure resistance measurements were carried out in a diamond-anvil cell apparatus (DACPPMS-ET225, Shanghai Anvilsource Material Technology Co., Ltd). The four-probe van der Pauw method was adopted to obtain the electrical resistance at high pressures. High-pressure XRD measurements were carried out on the BL15U1 beamline at Shanghai Synchrotron Radiation Facility with $\lambda = 0.6199$ Å. The in-situ pressure values inside the DAC were determined by the ruby fluorescence method [50] or the diamond Raman shift method [51], respectively.

**Theoretical Calculations.** The first-principles calculations were performed based on the density functional theories (DFT). We used the Vienna ab initio Simulation Package (VASP) [52] for electronic properties. To achieve consistency with experiments in lattice constants, we adopted non-local vdW-DF functional vdw-DF2 [53] for the exchange-correlation function. The ionic potentials are described in the framework of the projector augmented wave (PAW) method [54]. We also considered the strong correlation behavior of Mn atoms by adding onsite Coulomb interaction U (DFT+U) [55]. For superconducting properties, we used the QUANTUM ESPRESSO (QE) package [56]. We adopted PBE functional [57] and GBRV pseudopotentials [58] throughout our QE calculations. More details of our theoretical methods are described in Supplementary Note 2 (Supporting information).

## Acknowledgments

We thank Lili Zhang for her kind help during the HPXRD measurements at the Shanghai Synchrotron Radiation Facility (SSRF) and the support of the User Experiment Assist System of SSRF for *in-situ* high-pressure Raman spectroscopy measurements. The calculations were carried out using supercomputers at the High Performance Computing Center of the Collaborative Innovation Center of Advanced Microstructures, the high-performance supercomputing center of Nanjing University. This work was supported by the National Key R&D Program of China (No. 2022YFA1403201), the National Natural Science Foundation of China (Nos. 12061131001, 11927809, 12125404, 12204231, and 123B2055).

**Author contributions:** H.H.W. conceived and supervised the whole study. The $MnB_4$ samples were grown and characterized by Z.-N. X., and Y.-J. Z. The transport measurements at ambient and high pressures were done by Y.-J. Z., Z.-N. X., Y. L., Y. Y., and Q. L. with assistance from H.-H. W. High-pressure XRD measurements were performed by Q. L., Y.-J. Z., and Z.-N. X. The theoretical calculations and analyses were done by Q.L., T. H., Y. Z., and J. S. All authors discussed the results. H.-H. W. and Q. L. analyzed the experimental data and wrote the manuscript with the inputs from all co-authors.

**Competing interests:** The authors declare that they have no competing interests.

**Data and materials availability:** All data needed to evaluate the conclusions in the paper are present in the paper. Additional data related to this paper may be requested from the authors.

# Supplementary Information

## Superconductivity up to 14.2 K in MnB$_4$ under pressure

Zhe-Ning Xiang[†], Ying-Jie Zhang[†], Qing Lu[†], Qing Li*, Yiwen Li, Tianheng Huang, Yijie Zhu, Yongze Ye, Jian Sun*, and Hai-Hu Wen*

National Laboratory of Solid State Microstructures and Department of Physics, Collaborative Innovation Center of Advanced Microstructures, Nanjing University, Nanjing 210093, China

email: liqing1118@nju.edu.cn, jiansun@nju.edu.cn, hhwen@nju.edu.cn
[†] These authors contributed equally to this work.

## Supplementary Note 1

Single crystals of MnB$_4$ were prepared by modified solid-state reaction method using iodine as the agent. The stoichiometric mixtures of Mn (0.2541 g, Alfa Aesar, 99.99%) and B (0.2 g, damas-beta, 99.9%) powder was placed in silica ampoule with I2 (0.2 g, Aladdin) as mineralizer. The ampoule was heated to 1273 K for 14 days and quenched to room temperature. We obtained MnB$_4$ single crystals with the average sizes of about 0.1mm×0.05mm×0.05mm. Note that the residual powder besides the crystals was also characterized as the phase purity MnB$_4$ phase. To avoid the effect of iodine on the appearance of superconductivity, we grow MnB$_4$ polycrystalline through conventional solid state reaction method without the use of I$_2$. High purity Mn and B powder with a moral ratio of 1:4.5 were weighed, ground, and pressed into a pellet. The excess B is used to compensate for losses during the synthesis process. The pellets were loaded in alumina crucibles, sealed in quartz ampoules and, heated at 1393 K for 3 days.

We have performed high pressure resistance measurements on four samples in this study, among which run2 was carried out on $MnB_4$ single crystal, run1, run3 and run4 were carried out on $MnB_4$ polycrystalline. All measurements in different runs on $MnB_4$ show consistent results. The classification of these four samples is presented in the table below.

|  | Form of sample | Elements used for synthesis | Max. Pressure |
|---|---|---|---|
| run1 | polycrystalline | Mn, B, $I_2$ | 83 GPa |
| run2 | single crystal | Mn, B, $I_2$ | 86 GPa |
| run3 | polycrystalline | Mn, B | 85 GPa |
| run4 | polycrystalline | Mn, B | 150 GPa |

## Supplementary Note 2:

Throughout our calculations, the valence electron configurations are $2s^22p^1$ for B and $3p^63d^54s^2$ for Mn. In our VASP calculations, the criteria for the convergence of total energy and force are $10^{-6}$ eV and $10^{-3}$ eV/Å, respectively. The cutoff energy for the plane-wave basis set is 500 eV, with a sample of $2\pi \times 0.025$ Å$^{-1}$ and $2\pi \times 0.0125$ Å$^{-1}$ for Γ-centered $k$-point mesh in the Brillouin zone for self-consistency and density of state calculations, respectively. The U value was set as 3.5 eV empirically [1]. In our QE calculations, the criteria for the convergence of total energy and force are $10^{-7}$ Ry and $10^{-4}$ Ry/Bohr, respectively. The cutoff energy for the plane-wave basis set and charge densities are 60 and 600 Ry, respectively. The phonon spectrums are calculated using density functional perturbation theory (DFPT) as implemented in QE. The denser $k$- point mesh, coarser $k$-point mesh, and $q$-point mesh are 18×18×18, 9×9×9, and 3×3×3, respectively. Superconducting transition temperature $T_c$ is calculated using Allen-Dynes equations as follows [2]:

$$T_c = \frac{\omega_{log}}{1.2} exp\left[-\frac{1.04(1+\lambda)}{\lambda - \mu^*(1+0.62\lambda)}\right]$$

Where $\omega_{log}$, $\lambda$, and $\mu^*$ represent the logarithmic average of the phonon frequency, the electron-phonon coupling strength, and retarded Coulomb pseudopotential, respectively. Normally, $\mu^* = 0.1$ is acceptable. $\lambda$ is calculated as [2,3]:

$$\lambda = 2\int_0^{+\infty} \frac{\alpha^2 F(\omega)}{\omega} d\omega$$

Where $\alpha^2 F(\omega)$ is the Eliashberg spectral function:

$$\alpha^2 F(\omega) = \frac{1}{2\pi\hbar N(E_f)} \frac{1}{N_q} \sum_{qj} \frac{\gamma_{qj}}{\omega_{qj}} \delta(\omega - \omega_{qj})$$

Where $N_q$ is the number of $q$-points in generating the phonon mesh and $N(E_f)$ is the density of states at the Fermi level. $\omega_{qj}$ and $\gamma_{qj}$ are the phonon frequency and phonon linewidth of a certain phonon mode $j$ with momentum $q$, respectively. $\omega_{log}$ can be written as:

$$\omega_{log} = exp\left[\frac{\int_0^{+\infty} \frac{\alpha^2 F(\omega)}{\omega} log\,\omega\, d\omega}{\int_0^{+\infty} \frac{\alpha^2 F(\omega)}{\omega} d\omega}\right] = exp\left[\frac{2}{\lambda}\int_0^{+\infty} \frac{\alpha^2 F(\omega)}{\omega} log\,\omega\, d\omega\right]$$

**References**

[1] Tesch, R. & Kowalski, P. M. Hubbard $U$ parameters for transition metals from first principles. *Phys. Rev. B* **105**, 195153 (2022).

[2] Allen, P. B. & Dynes, R. C. Transition temperature of strong-coupled superconductors reanalyzed. *Phys. Rev. B* **12**, 905–922 (1975).

[3] Allen, P. B. Neutron Spectroscopy of Superconductors. *Phys. Rev. B* **6**, 2577–2579 (1972).

**Table S1** Summary of the superconducting transition temperature $T_c$ and the ratio of $\mu_0H_{c2}(0)/T_c$ for the reported B-containing superconductors.

| No. | Compound | Pressure (GPa) | $T_c$ (K) | $\mu_0H_{c2}(0)/T_c$ | Ref. |
|---|---|---|---|---|---|
| 1 | MgB$_2$ | AP | 39 | 0.64 | [1] |
| 2 | MoB$_2$ | 109.7 | 32.4 | 0.29 | [2] |
| 3 | YPd$_2$B$_2$C | AP | 23 | 0.44 | [3] |
| **4** | **MnB$_4$** | **150** | **14.2** | **0.94** | **This work** |
| 5 | LuRh$_4$B$_4$ | AP | 11.5 | 0.02 | [4] |
| 6 | LuRuB$_2$ | AP | 9.99 | 0.57 | [5] |
| 7 | YB$_6$ | AP | 7.55 | 0.04 | [6] |
| 8 | CrB$_2$ | 110.4 | 7.3 | 0.80 | [7] |
| 9 | Mo$_2$BC | AP | 7.1 | 0.39 | [8] |
| 10 | ZrB$_{12}$ | AP | 5.82 | 0.01 | [9] |
| 11 | Ba$_{0.67}$Pt$_3$B$_2$ | AP | 5.8 | 0.37 | [10] |
| 12 | Re$_3$B | AP | 5.1 | 0.69 | [11] |
| 13 | LuOs$_3$B$_2$ | AP | 4.67 | 0.43 | [12] |
| 14 | YB$_2$C$_2$ | AP | 3.6 | / | [13] |
| 15 | Ru$_7$B$_3$ | AP | 3.3 | 0.36 | [14] |
| 16 | FeB$_4$ | AP | 2.9 | 0.35 | [15] |

AP: ambient pressure.

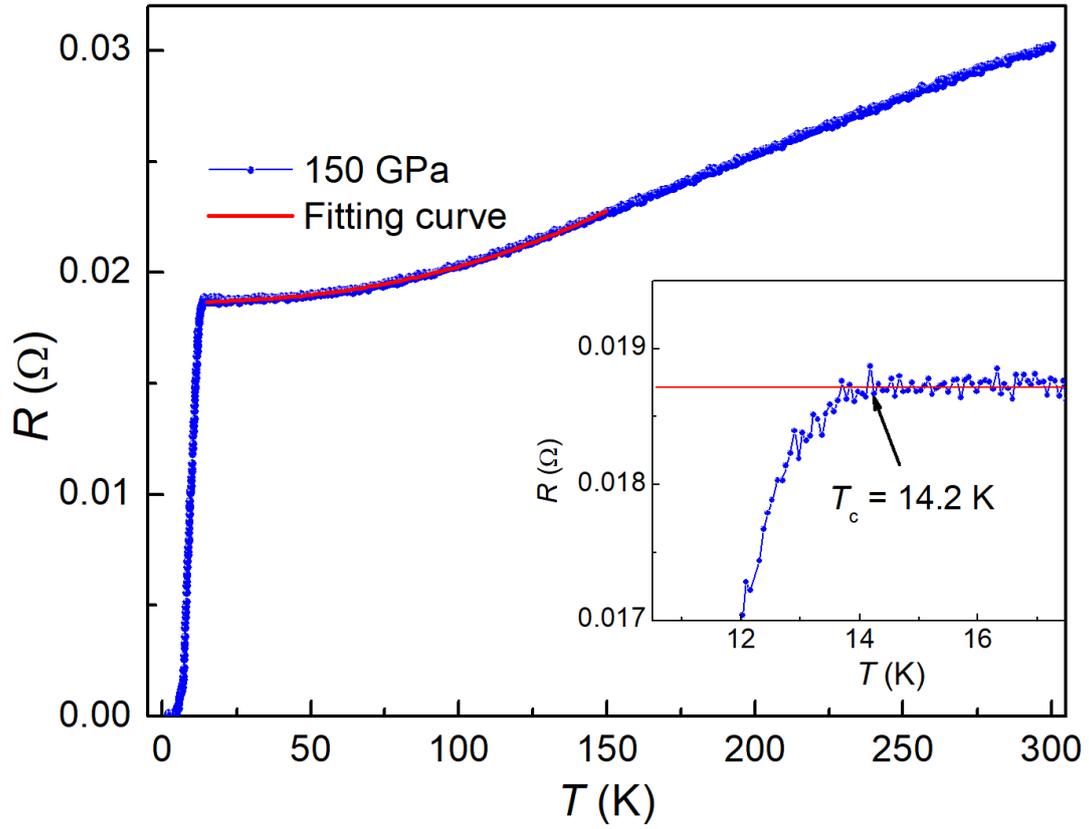

**Fig. S1** Temperature-dependent resistance for MnB$_4$ at 150 GPa. The solid red line is the fitting curve by using the formula $R(T) = R_0 + AT^n$, where $R_0$ is the residual resistance at zero temperature, $A$ and $n$ are the fitting parameters. The fitting gives the values of exponent $n$ is 2.3. Inset shows the highest $T_c$ observed in MnB$_4$.

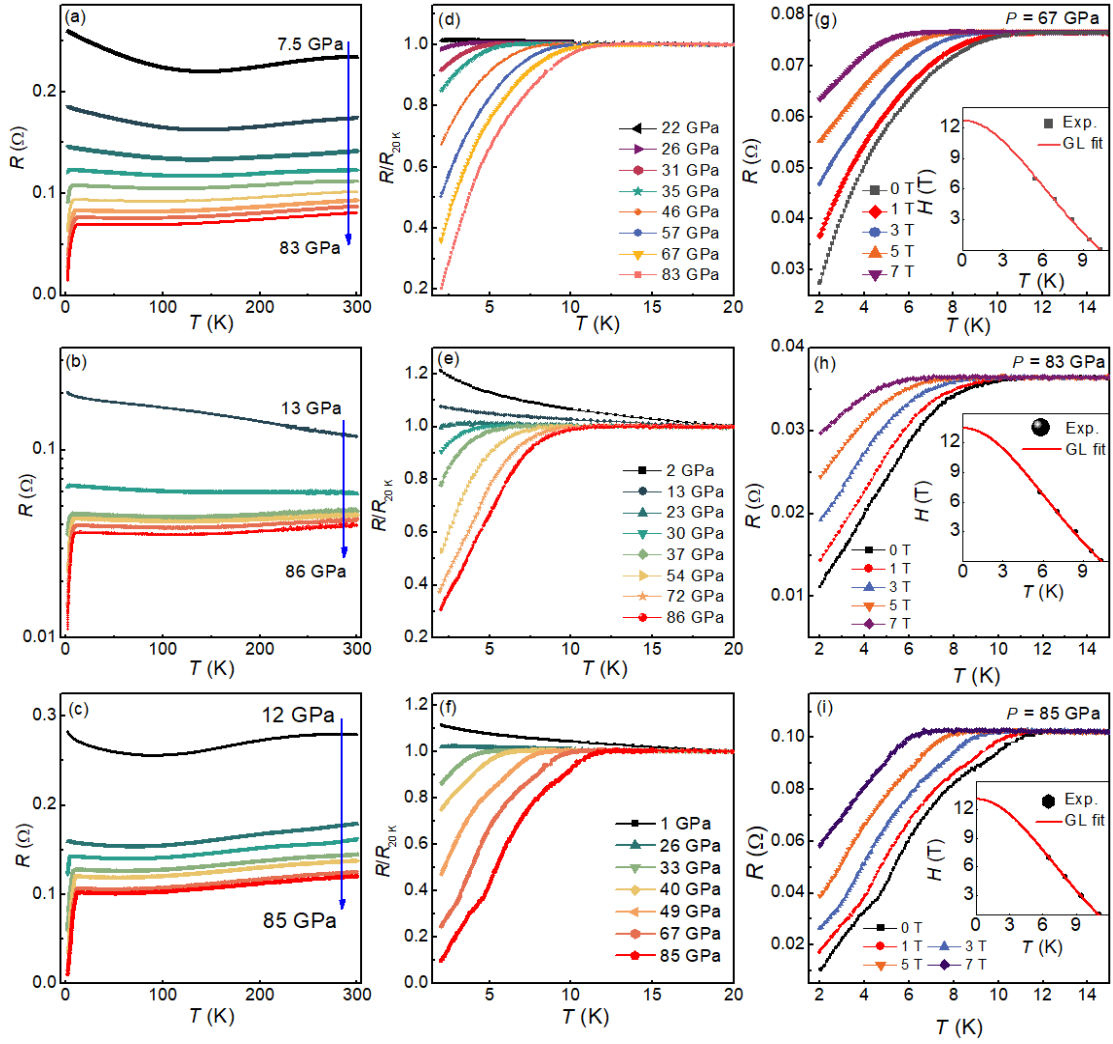

**Fig. S2** (a-c) Temperature-dependent resistance R-T of MnB$_4$ under various pressures for different runs. (d-f) Normalized R-T curves of MnB$_4$ from 2 to 20 K at selected pressures for different runs. (g-i) Temperature-dependent normalized resistance under various magnetic fields up to 7 T for different runs. Insets show the fitting curves by using the Ginzburg-Landau (G-L) equation $\mu_0 H_{c2}(T) = \mu_0 H_{c2}(0)[1-(T/T_c)^2]/[1+(T/T_c)^2]$.

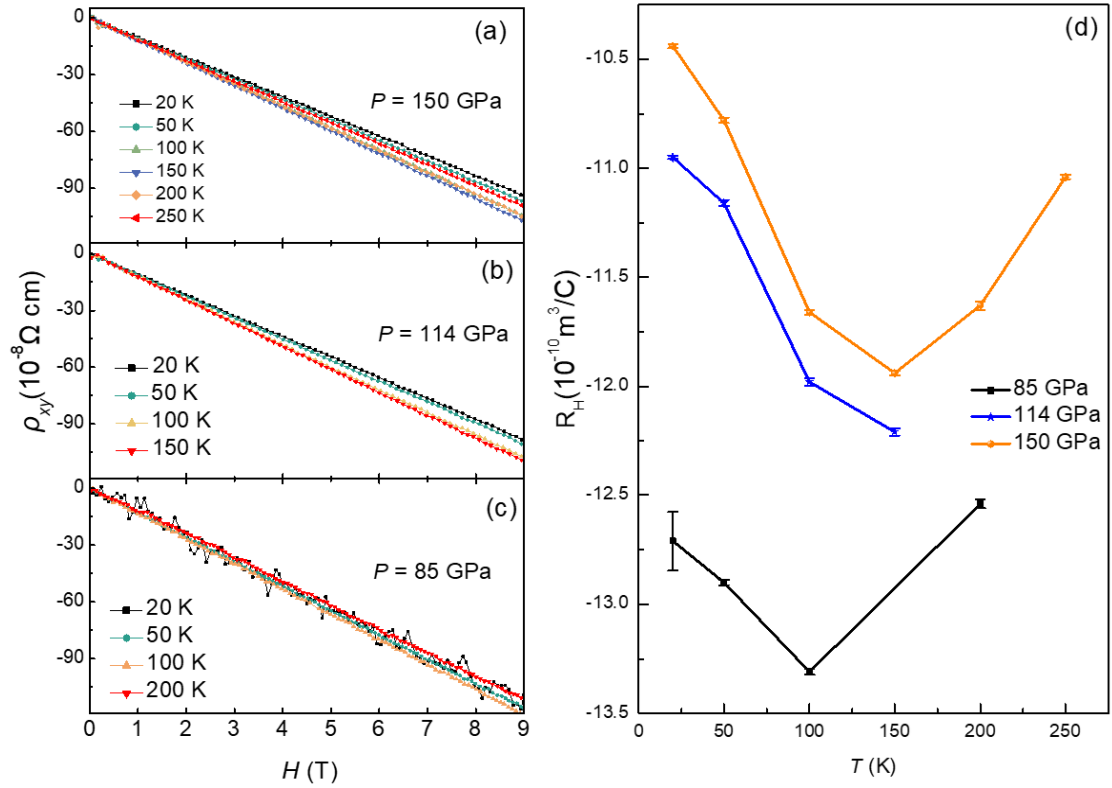

**Fig. S3** (a) Magnetic field dependence of the transverse resistivity $\rho_{xy}$ at different temperatures under different pressures. (b) The temperature dependence of Hall coefficient $R_H$. The error bar is derived from the scattering range of $\rho_{xy}$ data.

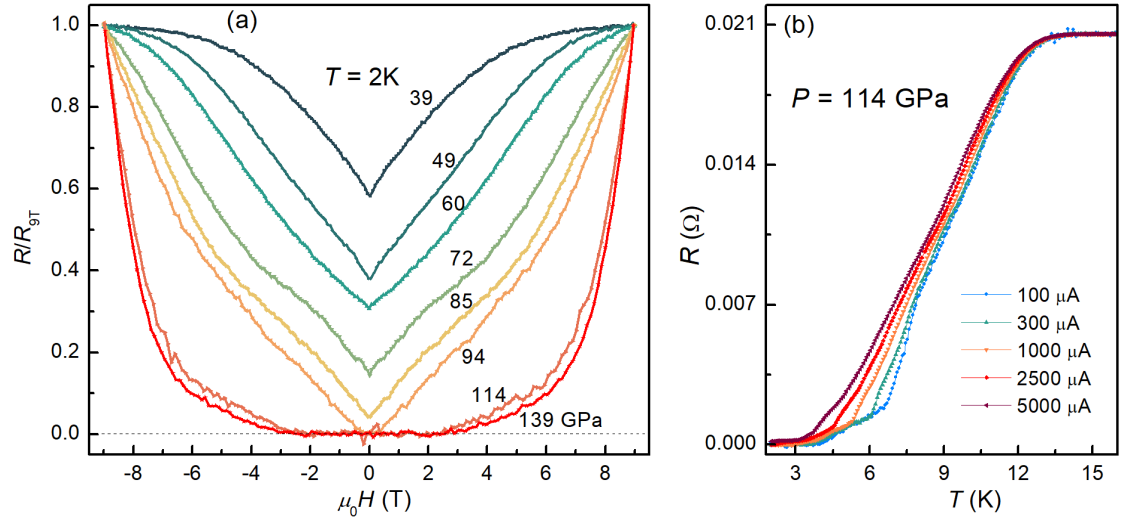

**Fig. S4** (a) Normalized magnetoresistance R/R$_{9T}$ of MnB$_4$ at 2 K under different pressures. The superconducting transition is gradually suppressed when the magnetic field is increasing. (b) Temperature-dependent resistance curve of MnB$_4$ at 114 GPa measured with different electric currents.

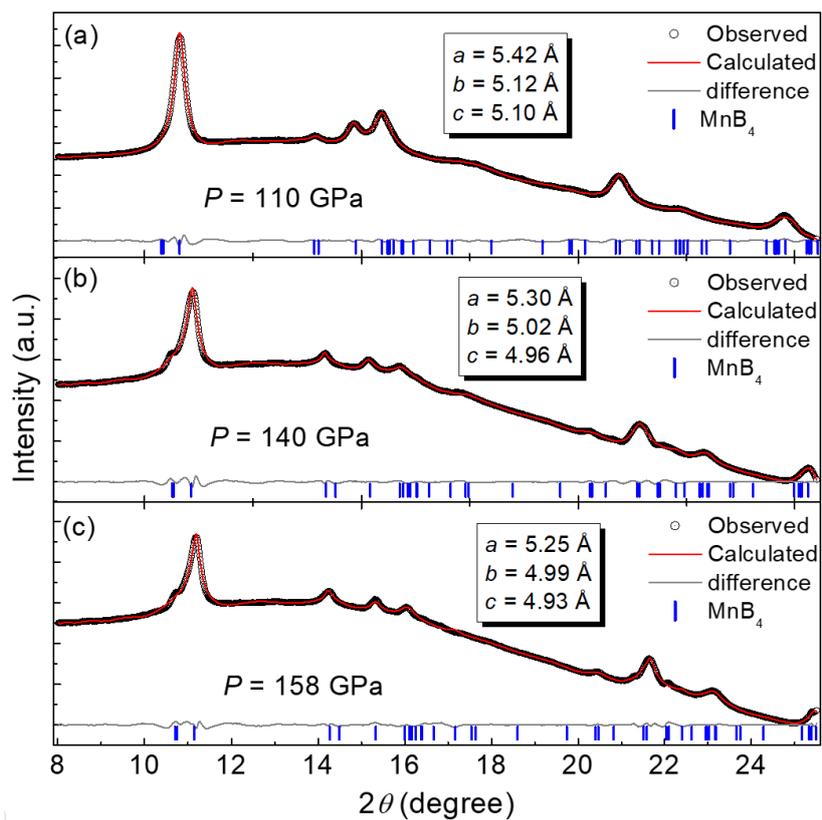

**Fig. S5** High-pressure X-ray diffraction patterns collected at (a) 110 GPa, (b)140 GPa, and (c)158 GPa for another $MnB_4$ sample. The black circles represent data of X-ray diffraction pattern and the red line represents the Rietveld fitting curve. The calculated lattice parameters at the corresponding pressures are listed in the inset.

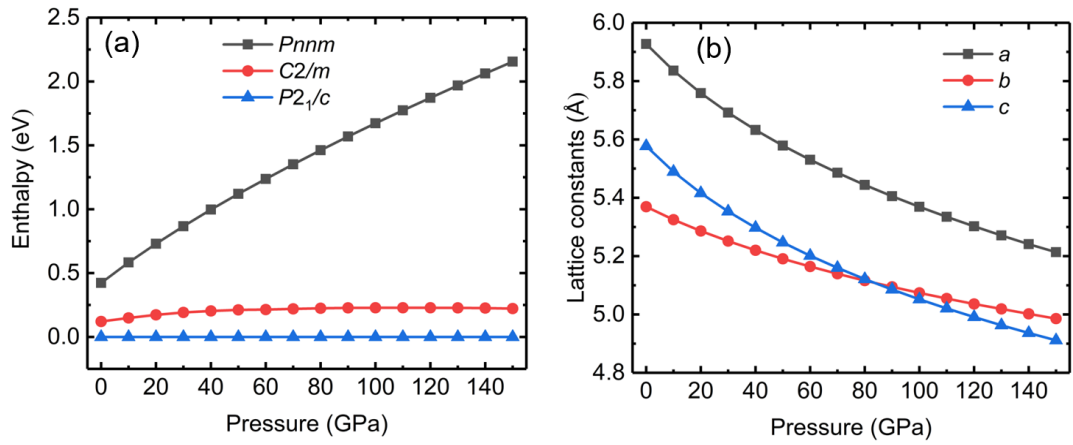

**Fig. S6** (a) Comparison of enthalpy curves between different possible phases of MnB$_4$. It is proved that the $P2_1/c$ phase is the most stable one. (b) Evolution of lattice constants of $P2_1/c$ phase under high pressures.

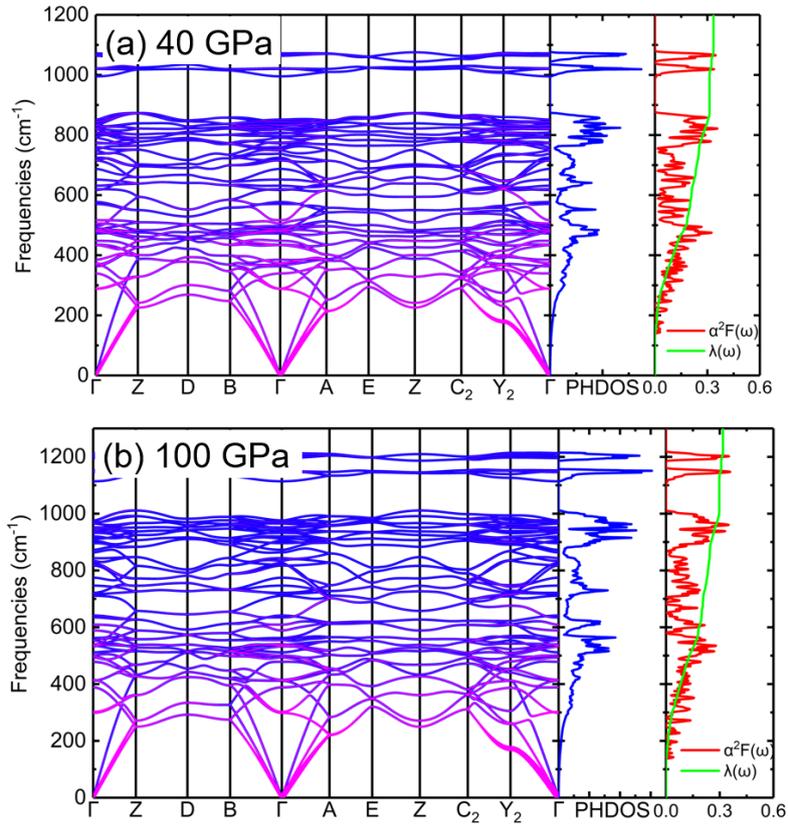

**Fig. S7** Calculated phonon spectrum, phonon density of states, Eliashberg function $\alpha^2F(\omega)$, and accumulated electron-phonon coupling strength $\lambda(\omega)$ at (a) 40 GPa, and (b) 100 GPa.

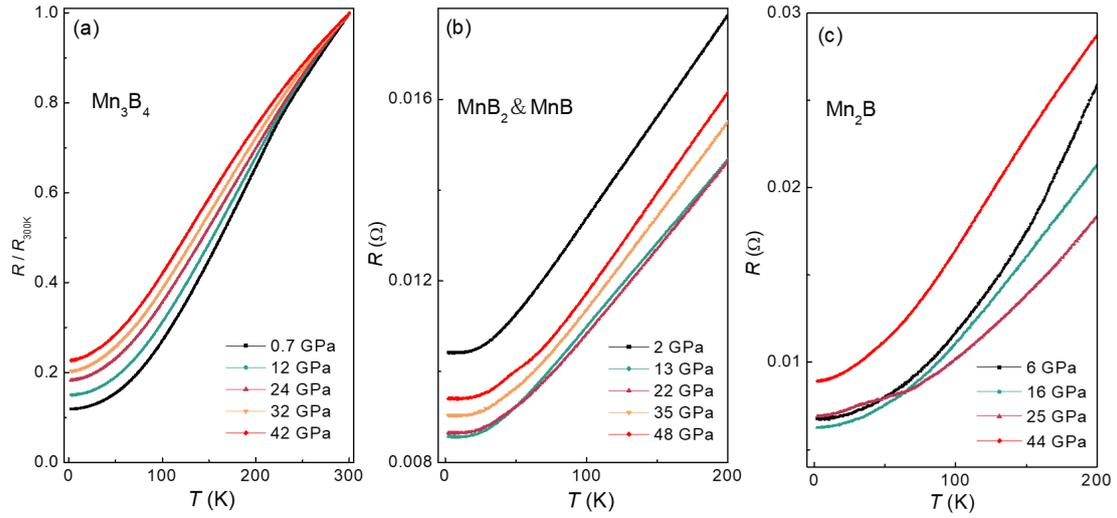

**Fig. S8** (a) Normalized temperature-dependent resistance of $Mn_3B_4$ under various pressures up to 42 GPa. (b) Temperature-dependent resistance up to 48 GPa for a compound contained both $MnB_2$ and MnB phase. (c) Temperature-dependent resistance of $Mn_2B$ under various pressures up to 44 GPa.